\documentclass[conference]{IEEEtran}
\IEEEoverridecommandlockouts
\usepackage{cite}
\usepackage{amsmath,amssymb,amsfonts}
\usepackage{algorithmic}
\usepackage{graphicx}
\usepackage{textcomp}
\usepackage{threeparttable}
\usepackage{booktabs}
\usepackage{xcolor}
\usepackage{multirow}
\usepackage{multicol}
\usepackage{subcaption}
\usepackage{xcolor}
\usepackage{makecell}
\definecolor{brightmaroon2}{RGB}{190,40,70}
\usepackage{tikz}
\usepackage{float}
\usepackage{tikz}
\usepackage{fancyhdr}
\usepackage{amsmath,amssymb}
\usetikzlibrary{arrows.meta, positioning, fit, calc}
\usepackage{array}
\newcolumntype{C}[1]{>{\centering\arraybackslash}p{#1}}
\newcolumntype{L}[1]{>{\raggedright\arraybackslash}p{#1}}
\usetikzlibrary{positioning,calc,fit,arrows.meta,backgrounds,decorations.pathreplacing}
\usepackage{tabularx}
\def\BibTeX{{\rm B\kern-.05em{\sc i\kern-.025em b}\kern-.08em
    T\kern-.1667em\lower.7ex\hbox{E}\kern-.125emX}}

\usepackage{orcidlink}
    
\begin{document}

\title{Diversity-Aware Ensemble of Quantum-Classical Models via Combinatorial Fusion \thanks{
The views expressed in this article are those of the authors and do not represent the views of Wells Fargo. This article is for informational purposes only. Nothing contained in this article should be construed as investment advice. Wells Fargo makes no express or implied warranties and expressly disclaims all legal, tax, and accounting implications related to this article.}
}





\author{
    \IEEEauthorblockN{
        Jingyan Xu\IEEEauthorrefmark{1}\orcidlink{0009-0006-0864-6035},
        Yuanhong Wu\IEEEauthorrefmark{1}\orcidlink{0009-0000-3226-9547}, 
        Samuel Yen-Chi Chen\IEEEauthorrefmark{2}\orcidlink{0000-0003-0114-4826},
        Sean Patten\IEEEauthorrefmark{1}\orcidlink{0009-0003-4850-8276},
        Christina Schweikert\IEEEauthorrefmark{3}\orcidlink{0009-0005-4721-8845}, \\
        and D. Frank Hsu\IEEEauthorrefmark{1}\orcidlink{0009-0007-9132-4367}\\
    }
    \IEEEauthorblockA{
        \IEEEauthorrefmark{1}Laboratory of Informatics and Data Mining, \\
        Department of Computer and Information Science, Fordham University, New York, NY, USA\\
    }
    \IEEEauthorblockA{
        \IEEEauthorrefmark{2}Wells Fargo, New York, NY, USA\\
    }
    \IEEEauthorblockA{
    \IEEEauthorrefmark{3}Division of Computer Science, Mathematics and Science, St. John's University, Queens, NY, USA\\
    }
}

\maketitle
\pagestyle{fancy}
\fancyhf{}
\fancyhead[L]{Published as a conference paper at 2026 IEEE Conference on Quantum Computing and Engineering (QCE26)}
\renewcommand{\headrulewidth}{0.4pt}

\begin{abstract}
Variational quantum circuits (VQCs) are highly sensitive to circuit architecture, yet most quantum machine learning pipelines select a single best design and discard the remaining candidates. This paper addresses whether the structural diversity across quantum circuit search spaces can be used to produce stronger predictions through model fusion. We apply Combinatorial Fusion Analysis (CFA), a model fusion framework that selects and combines multiple scoring systems using rank-score function (RSF) and cognitive diversity (CD), to a pool of 29 architecturally diverse classical models and quantum circuits. Using a diversity-aware model selection method, we identify subsets of maximally diverse models and combine them with both score and rank combination. Experiments on four binary classification benchmarks covering different domains demonstrate that CFA-based hybrid classical-quantum fusion consistently outperform the best individual model on every dataset considered with improvements up to 6.03\% in accuracy. When comparing with relevant literatures, the CFA framework further achieves superior performance across all four datasets.

\end{abstract}


\section{Introduction}

Quantum machine learning (QML) has emerged at the intersection of quantum computing and classical machine learning, offering a hybrid computational paradigm in which variational quantum circuits (VQCs) serve as trainable function approximators within classical optimization loops \cite{cerezo_variational_2021, bharti_noisy_2022}. Circuits built on VQCs have been developed for classification \cite{mitarai_quantum_2018, schuld_evaluating_2019, chen_end--end_2021}, time-series forecasting \cite{chen_quantum_2022, chen_learning_2024}, speech and language processing \cite{yang_decentralizing_2021, di_sipio_dawn_2022}, and reinforcement learning \cite{chen_variational_2020, chen_quantum_2023, lockwood_reinforcement_2020}, with several demonstrating competitive or superior performance relative to classical baselines under constrained parameter budgets.

A major barrier to wider use of quantum machine learning is circuit design. The performance of a variational quantum circuit depends heavily on its architecture. Choices of circuit architectures, the ansatz, such as encoding gates, entanglement patterns, and rotation layers all matter in a great deal. And the number of possible designs grows quickly as we add more qubits or increase depth. A circuit that works well for one task can fail on another. Designing these circuits takes specialized knowledge that many practitioners do not have \cite{chen_differentiable_2024}. Differentiable Quantum Architecture Search (DiffQAS) offers a way to automate this process. It assigns trainable weights to a pool of candidate circuit structures and uses gradients to identify strong architectures \cite{chen_differentiable_2024}. But DiffQAS and similar methods focus on selecting a single best design, while in this process, there may be some useful information being left behind spread across the other candidates.

Relying on a single model goes against a basic idea in classical machine learning, which is that a group of diverse models often performs better than any one model alone \cite{goos_ensemble_2000, zhou_ensemble_2012}. What makes an ensemble effective is not only the accuracy of each model, but also how different they are from each other. If their errors are uncorrelated and their decision boundaries complement one another, the combined result can be stronger \cite{kuncheva_combining_2004, brown_diversity_2005}. This idea has shaped decades of work in classical ML, including boosting, bagging, and stacking. In quantum machine learning, ensemble methods are less developed. Most existing approaches combine quantum circuits through simple averaging or ad hoc weighting, but they do not include a clear way to measure or use the diversity between models \cite{park_its--me_2025}.

Combinatorial Fusion Analysis (CFA) \cite{hsu_methods_2022, hsu_combinatorial_2006}, developed by Hsu, Chung, and Kristal, offers a general framework for analyzing and combining multiple scoring systems using score combination and rank combination. CFA characterizes each system through its rank-score function (RSF) and measures the dissimilarity between systems using a measure called cognitive diversity \cite{hurley_multi-layer_2021, hsu_cognitive_2019}. This framework provides both a basis foundation, which establishes the conditions under which fusion models can outperform individual models, and practical methods for selecting which systems to combine and how to combine. CFA has demonstrated consistent improvements across a wide range of domains, including drug discovery \cite{yang_consensus_2005, jiang_enhancing_2023}, text classification \cite{patten_enhancing_2025, xu_enhancing_2025, li_combination_2013}, bitcoin price prediction \cite{wu_bitcoin_2025}, value alignment \cite{wu_enhancing_2026}, cognitive neuroscience \cite{schweikert_detecting_2016}, and cybersecurity \cite{owusu_generalizable_2025}. The multi-layer extension of CFA on the Kemeny rank space further allows iterative improvement of combinations through expansion-reduction optimization \cite{hurley_multi-layer_2021, zhong_combining_2019}.



This paper applies CFA to quantum machine learning. Existing ensemble work in QML relies on simple averaging or ad hoc weights. To the authors' current knowledge, there is no prior method that uses a diversity measure to derive the selection and combination of variational quantum circuits. VQCs are a good target for this because circuits with different encoding axes, entanglement topologies, or variational gate sets act on data in very different ways. Even after training on the same dataset their predictions diverge in patterns that simple averaging tends to wash out. CFA offers us cognitive diversity (CD), a measure of how differently two scoring systems behave, and combination rules that use it. After treating each quantum circuit or classical baseline as an independent scoring system, we use CFA to select a diverse subset of those scoring systems and then combine their predictions. We hypothesize that this fusion outperforms the best individual model in the pool.

We conduct an extensive empirical evaluation across 4 binary classification datasets in different domains and sample sizes. Domains consist of medical diagnosis, signal processing, and cybersecurity, and dataset sizes ranging from small to thousands of samples. We also developed a selection method that not only prioritizes cognitive diversity but also takes performance of the circuits and models into account \cite{lin_feature_2007, hsu_cognitive_2019}. We then test several fusion strategies, including score combination, rank combination, and weighted versions of both.

The remainder of this paper is organized as follows. Section II reviews the relevant literature. Section III presents the datasets used in this study. Section IV explains the theoretical foundations of variational quantum circuits and quantum architecture search. Section V describes the classical models used in this hybrid model fusion experiments. Section VI details the combinatorial fusion analysis methodology. Section VII discusses the results and findings. Section VIII concludes with directions for future work.

\section{Related Work}

\subsection{Variational Quantum Circuits for Supervised Learning}

A variational quantum circuit (VQC) is a parameterized unitary optimized by classical gradient descent, with gradients computed analytically through the parameter-shift rule~\cite{peruzzo_variational_2014, mitarai_quantum_2018, schuld_evaluating_2019}. A VQC used for supervised learning typically has three stages: an encoding map that loads classical features into a quantum state, a sequence of trainable variational layers built from single-qubit rotations and entangling gates, and measurement of a set of observables. In hybrid quantum-classical models, this circuit is placed between classical input and output projections so that the entire pipeline can be trained end to end, an arrangement that has shown competitive performance against classical baselines under matched parameter budgets~\cite{chen_end--end_2021}. The choice of encoding strategy strongly affects what functions a VQC can represent \cite{sim_expressibility_2019}. Re-injecting the input at every variational layer, known as data re-uploading, extends the representable Fourier spectrum from degree~1 to degree~$L$ for $L$ layers~\cite{perez-salinas_data_2020}. VQCs have been studied across a range of supervised tasks beyond binary classification, including sequence modeling~\cite{chen_quantum_2022, chen_learning_2024}, vision and language tasks~\cite{yang_decentralizing_2021, di_sipio_dawn_2022}, and quantum kernel methods that embed data in Hilbert space for separation by a classical linear classifier~\cite{havlicek_supervised_2019}. A persistent obstacle to scaling VQCs is the barren plateau phenomenon, where gradients vanish exponentially with system size for deep or randomly initialized circuits~\cite{grant_initialization_2019, cerezo_variational_2021, larocca_barren_2025}, a difficulty compounded by the noise and depth constraints of current noisy intermediate-scale quantum (NISQ) hardware~\cite{preskill_quantum_2018, bharti_noisy_2022}.

\subsection{Quantum Architecture Search}


VQC performance is sensitive to circuit structure, and the space of plausible circuits grows too quickly with qubits and depth for manual design to keep up. Automated quantum architecture search (QAS)~\cite{martyniuk_quantum_2024} addresses this, with early methods using evolutionary search~\cite{altares-lopez_automatic_2021, ding_evolutionary_2022} or reinforcement learning~\cite{kuo_quantum_2021, sogabe_model-free_2022, wang_rnnpolicy_2024, rapp_reinforcement_2025} over discrete circuit spaces. Both scale poorly, as discrete spaces blow up combinatorially, which pushes evolutionary methods onto HPC-scale parallelism, and RL controllers are sample-inefficient and sensitive to reward design.

Differentiable QAS, such as DiffQAS~\cite{chen_differentiable_2024, zhang_differentiable_2022}, avoids these costs by relaxing the discrete architecture choice into a continuous mixture over candidate components, so that architecture weights and circuit parameters are learned jointly by gradient descent, in the spirit of DARTS~\cite{liu_darts_2019}. DiffQAS was originally developed for quantum reinforcement learning~\cite{chen_variational_2020, chen_quantum_2023, lockwood_reinforcement_2020}, but its design space over encoding gates, entanglement topologies, and initialization options is task-agnostic. Like other NAS-based methods, however, DiffQAS ultimately returns a single chosen architecture and discards every other candidate evaluated during the search. A natural question is whether the full pool of candidates, rather than just the winner, contains useful information that an ensemble could exploit.

\subsection{Ensemble Methods and Diversity}


The idea of combining classifiers to improve prediction predates modern machine learning by decades. Dietterich provides several reasons why ensembles work~\cite{goos_ensemble_2000}. First of all, averaging reduces the variance of individual predictions, but only when those predictions are not perfectly correlated. Diverse models contribute independent variance for averaging to cancel. Second, training multiple models explores more of the hypothesis space than a single optimizer would reach. A single training run is one pass of gradient descent. Different runs land at different local optima. Therefore, combining different runs samples a richer region of hypothesis space. Third, combinations of models is effectively averaging over this wider search, which is closer to the true target function than any single local optimum. The benefit vanishes when the models are identical.

\begin{table*}[htp]
\centering
\caption{\textbf{Summary of the four benchmark datasets} used in our paper. The majority-class ratio is the proportion of samples in the most frequent class.}
\label{tab:datasets}
\begin{tabular}{lrrp{1.8cm}p{8.5cm}}
\toprule
Dataset & \#Obs. & \#Feat. & Majority class ratio & Description \\
\midrule
Heart disease & 270 & 13 & 0.556 & Heart-related clinical data; predict presence vs absence of heart disease. \\
Ionosphere & 351 & 34 & 0.641 & Radar returns; predict good vs bad ionospheric signals. \\
Pima Diabetes & 768 & 8 & 0.651 & Medical measurements; predict diabetes outcome (0/1). \\
Spambase & 4601 & 57 & 0.606 & Email contents; classify spam vs non-spam. \\
\bottomrule
\end{tabular}
\end{table*}

Standard methods such as bagging~\cite{breiman_bagging_1996}, boosting~\cite{freund_decision-theoretic_1997}, random forests~\cite{breiman_random_2001}, and stacking~\cite{wolpert_stacked_1992} produce diverse base learners through randomized training, sequential error correction, or architectural variation. Common ways to aggregate predictions include averaging, voting, or learned weights. The ambiguity decomposition shows that ensemble error equals the average individual error minus the average pairwise diversity~\cite{krogh_neural_1994, kuncheva_combining_2004}. Therefore, it's been long established that diversity drives ensemble performance. Yet over a dozen diversity measures have been proposed~\cite{kuncheva_measures_2003, brown_diversity_2005}, none dominates, and standard ensembles do not measure or optimize diversity during model selection. The same gap exists, in a more pronounced form, in quantum machine learning, where ensemble approaches remain limited \cite{park_its--me_2025} and typically rely on simple averaging of output probabilities without any formal diversity criterion.

\subsection{Combinatorial Fusion Analysis}

Combinatorial Fusion Analysis (CFA) is a framework for combining multiple scoring systems~\cite{hsu_methods_2022, hsu_combinatorial_2006}. It has two central components: (1) the rank-score function (RSF), which maps each system's rank positions to its score values, and (2) cognitive diversity~\cite{hsu_cognitive_2019}, which measures dissimilarity between two systems as the $\ell_2$ distance between their rank-score functions. Unlike Pearson correlation, Spearman~$\rho$, or the prediction-disagreement statistics common in ensemble learning, cognitive diversity is a measurement of dissimilarity. CD captures how differently two models distribute discrimination across the ranked population, not just how correlated their outputs are. CFA establishes two joint criteria for effective combination~\cite{hsu_methods_2022, lin_feature_2007}: each individual scoring system must have relatively good performance, and the systems must be cognitively diverse from each other. A further theorem shows that under sufficiently high cognitive diversity, rank combination outperforms score combination~\cite{hsu_methods_2022}.

CFA has been applied to virtual screening for drug discovery~\cite{yang_consensus_2005, jiang_enhancing_2023}, protein structure prediction~\cite{lin_feature_2007}, intrusion detection~\cite{owusu_generalizable_2025}, and financial forecasting~\cite{wu_bitcoin_2025}. The multi-layer combinatorial fusion extension~\cite{hurley_multi-layer_2021} formalizes iterative selection and combination on the Kemeny rank space~$H_n$ and supports tied rankings natively. CFA has not previously been applied to quantum machine learning, and no prior work has used cognitive diversity to guide the selection and combination of variational quantum circuits.

\subsection{CFA versus Standard Ensemble Methods}

CFA differs from standard ensembles on three points. The first is the diversity measure itself. Cognitive diversity is a geometric distance between rank-score functions, not a statistic over disagreeing predictions. It captures structural differences in scoring behavior that are overlooked by correlation-based measures. The second is where the combination happens. Standard ensembles commit to one space, scores for soft voting or ranks for hard voting, without much theoretical guidance for the choice. CFA operates in both spaces and reveals which is preferable for a given model pool: under high cognitive diversity, rank combination outperforms score combination \cite{hsu_methods_2022}. The third concerns when models are filtered. Standard ensembles combine everything and let the aggregator manage redundancy. CFA filters first, jointly optimizing individual quality and pairwise dissimilarity before any combination is performed. The fit with quantum machine learning is natural, since VQC architectures already differ in encoding axis, entanglement topology, and gate structure. Naturally, cognitive diversity is well-suited to measure this kind of variation.

\section{Datasets}



We evaluate our framework on four benchmark binary classification datasets, summarized in Table~\ref{tab:datasets}: \emph{Heart disease}, \emph{Ionosphere}, \emph{Pima Indians Diabetes}, and \emph{Spambase}. They span medical diagnosis, signal processing, and cybersecurity, with 270--4601 samples and 8--57 features per dataset \cite{caro_generalization_2022}.

The Heart disease dataset is a collection of medical records designed for predicting the presence and severity of cardiovascular disease. The Ionosphere dataset classifies radar returns from the ionosphere as either structure detected or no structure. The Pima Indians Diabetes dataset is used to diagnostically predict whether an Pima Indian patient has diabetes. The Spambase dataset is used to classify emails as spam or non-spam. 

All datasets pass through a unified preprocessing pipeline. First, labels are encoded as $\{0,1\}$; categorical columns are ordinal-encoded. Then we handle missing values by using mean imputation. Furthermore, we use the training set to z-score normalize the features. We then apply a stratified 70-15-15 train-validation-test split to preserve class proportions.

\section{Quantum circuits architecture}

This section introduces the quantum circuits that form the model pool evaluated in our combinatorial fusion experiments. Our quantum architecture search framework is a custom variant of DiffQAS-QRL \cite{chen_differentiable_2024} that enumerates a structured design space. Training every configuration in this space gives us the 12 circuits used as base models. In addition, we use another 8 carefully curated circuits that are architecturally diverse. Both sets of circuits are wrapped in a unified classification pipeline which consists of training, evaluation, and comparison process within the quantum fusion framework.

\begin{figure}[htbp]
\centering
\begingroup
\resizebox{0.75\linewidth}{!}{%
\begin{tikzpicture}[
    >=Stealth,
    every node/.style={font=\small},
    vqcblock/.style={draw, rectangle, minimum height=1.0cm,
                     minimum width=5.0cm, align=center, rounded corners=3pt},
    vqcdashedbox/.style={draw, dashed, line width=0.7pt,
                         inner xsep=0.9cm, inner ysep=0.35cm}
]

\node (x) at (0,8.2) {$x \in \mathbb{R}^d$};
\node[vqcblock, fill=orange!15] (win) at (0,7.0)
    {$W_{\mathrm{in}} + \tanh$};

\node[vqcblock, fill=blue!12] (encode) at (0,4.8)
    {Encode $R(x)$};
\node[vqcblock, fill=violet!12] (rot) at (0,2.8)
    {$R(\theta) + \mathrm{Entangling}$};
\node[vqcblock, fill=green!12] (measure) at (0,0.8)
    {Measure $\langle Z_0\rangle,\ldots, \langle Z_{n-1}\rangle$};

\node[vqcblock, fill=orange!15] (wout) at (0,-1.2)
    {$\tanh + W_{\mathrm{out}} + \mathrm{softmax}$};
\node (yhat) at (0,-2.4) {$\hat{y}$};

\draw[->, thick] (x) -- (win);
\draw[->, thick] (win) -- (encode);
\draw[->, thick] (encode) -- (rot);
\draw[->, thick] (rot) -- (measure);
\draw[->, thick] (measure) -- (wout);
\draw[->, thick] (wout) -- (yhat);

\coordinate (rTop) at ($(encode.east)+(1.2,0)$);
\coordinate (rDown) at ($(rot.east)+(1.2,0)$);
\coordinate (rMidTop) at ($(rTop)!0.38!(rDown)$);
\coordinate (rMidBot) at ($(rTop)!0.72!(rDown)$);

\draw[thick] (encode.east) -- (rTop);
\draw[thick] (rTop) -- (rMidTop);
\draw[thick] (rMidBot) -- (rDown);
\draw[->, thick] (rDown) -- (rot.east);

\node[align=left, anchor=west] at ($(rMidTop)!0.5!(rMidBot)+(-1,0)$)
    {re-upload $x$\\ \hspace{0.68cm}$\times L$};

\node[vqcdashedbox, fit=(encode)(rot)(measure)(rTop)(rDown)] (qbox) {};

\node[anchor=west] at (4.9,7.0) {\textbf{classical}};
\node[anchor=west] at (4.9,3.2) {\textbf{quantum}};
\node[anchor=west] at (4.9,-1.2) {\textbf{classical}};

\end{tikzpicture}}
\endgroup
\caption{\textbf{VQC architecture.}}
\label{fig:vqc_architecture}
\end{figure}
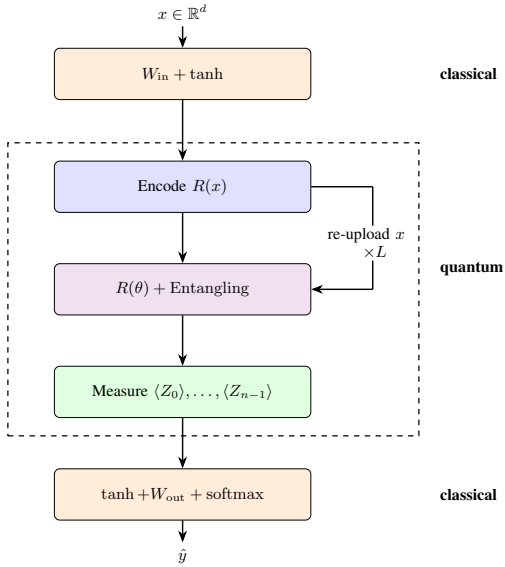

\subsection{Variational Quantum Circuits: a general structure}

A variational quantum circuit (VQC) is a parameterized unitary transformation $U(\mathbf{x}, \boldsymbol{\theta})$ acting on a register of $n$ qubits, where $\mathbf{x} \in \mathbb{R}^n$ is classical input data and $\boldsymbol{\theta}$ is a set of trainable rotation angles optimized by gradient descent.

Classical features are loaded into the quantum state by single-qubit rotations, $L$ variational layers then interleave parameterized rotations with two-qubit entangling gates, and the circuit is read out by measuring Pauli-$Z$ expectation values $\langle Z_0\rangle, \ldots, \langle Z_{n-1}\rangle \in [-1,1]^n$. We simulate all circuits on PennyLane's backend through the PyTorch interface, with parameter broadcasting enabled for batched execution.

\subsection{Data re-uploading}



All 12 search-space configurations use data re-uploading~\cite{perez-salinas_data_2020}. Each layer $k$ has its own block of trainable parameters $\boldsymbol{\theta}_k$, and the input projection is added to the rotation angles at every layer instead of only at layer~0. The number of variational layers is set to $\lceil n_{\text{features}} / n_{\text{qubits}} \rceil$, so circuits that compress more features into each qubit go deeper. The exact three-phase structure used by the VQCs in this work is shown in Figure \ref{fig:vqc_architecture}.

\subsection{Circuit Architecture Search Space}

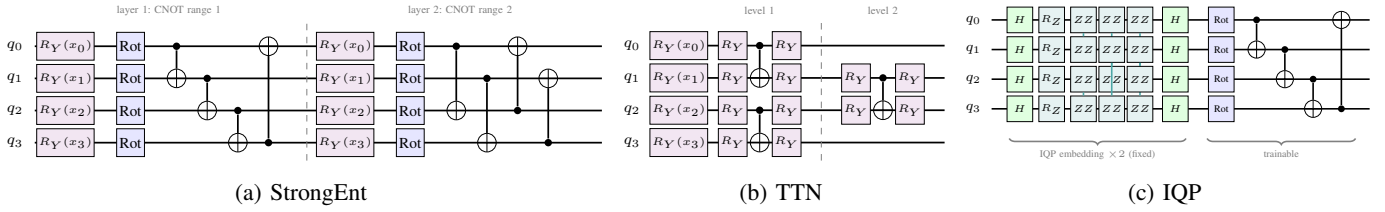
\begin{figure*}[htbp]
\begingroup
\tikzset{
  plc/qwire/.style  = {thick},
  plc/gate/.style   = {draw, fill=blue!10, minimum size=0.45cm,
                   font=\scriptsize, inner sep=1pt},
  plc/gateg/.style  = {draw, fill=green!12, minimum size=0.45cm,
                   font=\scriptsize, inner sep=1pt},
  plc/gatery/.style = {draw, fill=violet!10, minimum size=0.45cm,
                   font=\scriptsize, inner sep=1pt},
  plc/gaterz/.style = {draw, fill=teal!10, minimum size=0.45cm,
                   font=\scriptsize, inner sep=1pt},
  plc/ctrl/.style   = {fill=black, circle, inner sep=1.2pt},
  plc/targ/.style   = {draw, circle, inner sep=0pt, minimum size=0.3cm,
                   path picture={
                     \draw (path picture bounding box.north)
                        -- (path picture bounding box.south)
                           (path picture bounding box.west)
                        -- (path picture bounding box.east);
                   }},
  plc/qlabel/.style = {font=\scriptsize, anchor=east},
}
\def\gstep{0.62}
\def\qsep{0.52}
\centering

\begin{subfigure}[b]{0.44\textwidth}
\centering
\resizebox{\linewidth}{!}{%
\begin{tikzpicture}[x=\gstep cm, y=-\qsep cm]
  \foreach \q in {0,...,3}{
    \draw[plc/qwire] (-0.3, \q) -- (14.5, \q);
    \node[plc/qlabel] at (-0.4, \q) {$q_{\q}$};
  }
  \foreach \q in {0,...,3}{\node[plc/gatery] at (0.5, \q) {\tiny$R_Y(x_{\q})$};}
  \foreach \q in {0,...,3}{\node[plc/gate] at (2.2, \q) {Rot};}
  \foreach \c/\t/\xp in {0/1/3.4, 1/2/4.2, 2/3/5.0}{
    \node[plc/ctrl] (c\c\t) at (\xp, \c) {};
    \node[plc/targ] (t\c\t) at (\xp, \t) {};
    \draw[thick] (c\c\t) -- (t\c\t);
  }
  \node[plc/ctrl] (c30) at (5.8, 3) {};
  \node[plc/targ] (t30) at (5.8, 0) {};
  \draw[thick] (c30) -- (t30);
  \draw[thin, gray, dashed] (6.8, -0.7) -- (6.8, 3.7);
  \foreach \q in {0,...,3}{\node[plc/gatery] at (7.8, \q) {\tiny$R_Y(x_{\q})$};}
  \foreach \q in {0,...,3}{\node[plc/gate] at (9.5, \q) {Rot};}
  \node[plc/ctrl] (a0) at (10.7, 0) {}; \node[plc/targ] (b0) at (10.7, 2) {};
  \draw[thick] (a0) -- (b0);
  \node[plc/ctrl] (a1) at (11.5, 1) {}; \node[plc/targ] (b1) at (11.5, 3) {};
  \draw[thick] (a1) -- (b1);
  \node[plc/ctrl] (a2) at (12.3, 2) {}; \node[plc/targ] (b2) at (12.3, 0) {};
  \draw[thick] (a2) -- (b2);
  \node[plc/ctrl] (a3) at (13.1, 3) {}; \node[plc/targ] (b3) at (13.1, 1) {};
  \draw[thick] (a3) -- (b3);
  \node[font=\tiny, gray] at (3.2, -1.1) {layer 1: CNOT range 1};
  \node[font=\tiny, gray] at (10.8, -1.1) {layer 2: CNOT range 2};
\end{tikzpicture}}
\caption{StrongEnt}
\label{fig:circ_strongent}
\end{subfigure}%
\hfill
\begin{subfigure}[b]{0.24\textwidth}
\centering
\resizebox{\linewidth}{!}{%
\begin{tikzpicture}[x=\gstep cm, y=-\qsep cm]
  \foreach \q in {0,...,3}{
    \draw[plc/qwire] (-0.3, \q) -- (7.3, \q);
    \node[plc/qlabel] at (-0.4, \q) {$q_{\q}$};
  }
  \foreach \q in {0,...,3}{\node[plc/gatery] at (0.4, \q) {\tiny$R_Y(x_{\q})$};}
  \node[plc/gatery] at (1.8, 0) {\tiny$R_Y$};
  \node[plc/gatery] at (1.8, 1) {\tiny$R_Y$};
  \node[plc/ctrl] (c01) at (2.5, 0) {};
  \node[plc/targ] (t01) at (2.5, 1) {};
  \draw[thick] (c01) -- (t01);
  \node[plc/gatery] at (3.2, 0) {\tiny$R_Y$};
  \node[plc/gatery] at (3.2, 1) {\tiny$R_Y$};
  \node[plc/gatery] at (1.8, 2) {\tiny$R_Y$};
  \node[plc/gatery] at (1.8, 3) {\tiny$R_Y$};
  \node[plc/ctrl] (c23) at (2.5, 2) {};
  \node[plc/targ] (t23) at (2.5, 3) {};
  \draw[thick] (c23) -- (t23);
  \node[plc/gatery] at (3.2, 2) {\tiny$R_Y$};
  \node[plc/gatery] at (3.2, 3) {\tiny$R_Y$};
  \node[plc/gatery] at (5.0, 1) {\tiny$R_Y$};
  \node[plc/gatery] at (5.0, 2) {\tiny$R_Y$};
  \node[plc/ctrl] (c12) at (5.7, 1) {};
  \node[plc/targ] (t12) at (5.7, 2) {};
  \draw[thick] (c12) -- (t12);
  \node[plc/gatery] at (6.4, 1) {\tiny$R_Y$};
  \node[plc/gatery] at (6.4, 2) {\tiny$R_Y$};
  \draw[thin, gray, dashed] (4.1, -0.7) -- (4.1, 3.7);
  \node[font=\tiny, gray] at (2.5, -1.1) {level 1};
  \node[font=\tiny, gray] at (5.7, -1.1) {level 2};
\end{tikzpicture}}
\caption{TTN}
\label{fig:circ_ttn}
\end{subfigure}%
\hfill
\begin{subfigure}[b]{0.30\textwidth}
\centering
\resizebox{\linewidth}{!}{%
\begin{tikzpicture}[x=\gstep cm, y=-\qsep cm]
  \foreach \q in {0,...,3}{
    \draw[plc/qwire] (-0.3, \q) -- (10.4, \q);
    \node[plc/qlabel] at (-0.4, \q) {$q_{\q}$};
  }
  \foreach \q in {0,...,3}{\node[plc/gateg] at (0.5, \q) {\tiny$H$};}
  \foreach \q in {0,...,3}{\node[plc/gaterz] at (1.4, \q) {\tiny$R_Z$};}
  \foreach \a/\b/\xp in {0/1/2.3, 2/3/2.3, 0/2/3.1, 1/3/3.1, 0/3/3.9, 1/2/3.9}{
    \node[plc/gaterz] (za\a\b) at (\xp, \a) {\tiny$ZZ$};
    \node[plc/gaterz] (zb\a\b) at (\xp, \b) {\tiny$ZZ$};
    \draw[thick, teal!70] (za\a\b) -- (zb\a\b);
  }
  \foreach \q in {0,...,3}{\node[plc/gateg] at (4.9, \q) {\tiny$H$};}
  \draw[decorate, decoration={brace, amplitude=3pt, mirror},thick, gray]
    (0.2, 4.0) -- node[below=2pt, font=\tiny, gray]
    {IQP embedding $\times 2$ (fixed)} (5.2, 4.0);
  \foreach \q in {0,...,3}{\node[plc/gate] at (6.2, \q) {\tiny Rot};}
  \node[plc/ctrl] (e0) at (7.2, 0) {}; \node[plc/targ] (f0) at (7.2, 1) {};
  \draw[thick] (e0) -- (f0);
  \node[plc/ctrl] (e1) at (8.0, 1) {}; \node[plc/targ] (f1) at (8.0, 2) {};
  \draw[thick] (e1) -- (f1);
  \node[plc/ctrl] (e2) at (8.8, 2) {}; \node[plc/targ] (f2) at (8.8, 3) {};
  \draw[thick] (e2) -- (f2);
  \node[plc/ctrl] (e3) at (9.6, 3) {}; \node[plc/targ] (f3) at (9.6, 0) {};
  \draw[thick] (e3) -- (f3);
  \draw[decorate, decoration={brace, amplitude=3pt, mirror},thick, gray]
    (5.8, 4.0) -- node[below=2pt, font=\tiny, gray]
    {trainable} (10.0, 4.0);
\end{tikzpicture}}
\caption{IQP}
\label{fig:circ_iqp}
\end{subfigure}

\caption{\textbf{Representative circuit architectures} on $n=4$ qubits, illustrating three design philosophies in the 8-circuit pool. (a) StrongEnt; (b) TTN; (c) IQP. The remaining five circuits (BasicEnt, TwoDesign, Random, QAOA, MPS) follow one of these three philosophies and are described in the text.}
\label{fig:pennylane_circuits}
\endgroup
\end{figure*}

The original DifferentiableQAS-QRL framework treats quantum circuit design as a search problem \cite{chen_differentiable_2024}. Instead of choosing one hand-built ansatz, it looks at all possible circuits you can create from a fixed set of design options across four independent dimensions. It then learns which designs to select or combine using differentiable optimization. In this work, we use the framework by building and testing every circuit in the search space as its own fixed classifier. 


  

Furthermore, we extends the original DiffQAS search space framework by replacing both the variational rotation gate and the entangling gate with more expressive parameterized alternatives \cite{chen_differentiable_2024}. Where the original framework uses a single-axis rotation (RX, RY, or RZ) for the variational layer and a fixed CNOT for entanglement, our new set of circuits uses a full $\text{Rot}(\phi, \theta, \omega)$ gate for every variational qubit and a parameterized $\text{CRZ}(\gamma)$ for every entangling pair. This change roughly triples the number of variational parameters per layer, from one angle per qubit to three, and replaces the rigid CNOT connectivity with a learnable interaction strength. Because the variational rotation axis is no longer a design choice (it is always $\text{Rot}$), the total number of configurations changed from the original 36 to 12 in total.

\subsubsection{The Three-Component Search Space}

Each of the 12 circuit configuration is determined by one choice from each of the following three components:

\textbf{Hadamard Initialization} —
\begin{itemize}
    \item H: A Hadamard gate is applied to every qubit before encoding, placing the system in an equal superposition $\frac{1}{\sqrt{2^n}}\sum_{j=0}^{2^n-1}|j\rangle$.
    \item noH: No Hadamard layer is applied. Each qubit begins in the standard computational state $|0\rangle^{\otimes n}$.
\end{itemize}

\textbf{Encoding Rotation} —
\begin{itemize}
    \item RX, RY, or RZ: The projected input features are encoded using the corresponding single-qubit rotation gate, identical in role to V1. The axis choice determines how input information is oriented in the Bloch sphere prior to the variational and entangling operations.
\end{itemize}

\textbf{Entangling Topology} —
\begin{itemize}
    \item Standard (std): $\text{CRZ}(\gamma)$ gates are applied in a brick-wall pattern — first to even-indexed pairs $(0,1), (2,3), \ldots$, then to odd-indexed pairs $(1,2), (3,4), \ldots$ — matching the spatial structure of the V1 standard topology but replacing the fixed CNOT with a learnable controlled-Z rotation.
    \item Cycle (cyc): $\text{CRZ}(\gamma)$ gates are arranged in a ring, with $\text{CRZ}(i,\,(i+1) \bmod n)$ connecting each qubit to the next in a closed loop.
\end{itemize}

\subsubsection{Circuit Structure}

Each new configuration shares the same outer architecture as the original framework:
$$\text{Input} \xrightarrow{\text{Linear}} \tanh \xrightarrow{\text{VQC}} \tanh \xrightarrow{\text{Linear}} \text{logits}$$
Within each VQC layer, the operations follow the order:
$$[\text{optional H}] \;\rightarrow\; \text{encode}(x) \;\rightarrow\; \text{Rot}(\phi, \theta, \omega) \;\rightarrow\; \text{CRZ}(\gamma)$$

The full design space contains $2 \times 3 \times 2 = 12$ configurations, each named using the pattern \{H$|$noH\}\_\{RX$|$RY$|$RZ\}\_\{std$|$cyc\}, for example \texttt{H\_RY\_std} or \texttt{noH\_RX\_cyc}. Every configuration is trained independently and contributes its own prediction scores as a separate scoring system in the CFA pool.

\subsection{Curated Quantum Circuits}

In addition to the 12 circuits derived from the search space, we also added 8 curated quantum circuits as our base models. The 12 search-space circuits are architecturally distinct from the following circuits which we will explain in detail. The 12 search-space circuits use a uniform template: encoding $(RX/RY/RZ)$ to parameterized CRZ entangling to Rot (ZYZ Euler) variational, across all 12 configurations. Meanwhile, the 8 curated circuits use completely different gate primitives.

Each of the eight curated circuits are architecturally distinct built. Each circuit uses the same outwrapper as the 12 search-space circuits. They each have a linear input projection from $d$ features to $n$ qubits passed through a $\tan h$ activation, followed by the variational quantum circuit, a second $\tan h$ activation on the circuit outpus, and a linear output projection to two logits for binary classification. All trainable parameters are initialized from $\mathcal{N}(0,1) \cdot \pi/2$ (except QAOA, which uses $\pi/4$), and all circuits measure Pauli-$Z$ expectations on every qubit. Figure \ref{fig:pennylane_circuits} illustrates three representative circuits, with each reflects different design philosophies in our curated circuit pool. 

\subsubsection{StrongEnt (StronglyEntanglingLayers)}



Each layer of StrongEnt applies a general $\mathrm{Rot}(\phi,\theta,\omega)$ rotation (three Euler angles) on every qubit, followed by a CNOT ladder whose control--target range increments each layer and cycles modulo~$n$~\cite{schuld_circuit-centric_2020}. The variable-range pattern ensures all qubit pairs become entangled over successive layers.

\subsubsection{BasicEnt (BasicEntanglerLayers)}



BasicEnt is the most parameter-efficient circuit among the curated circuits we are using. It consists of a single $R_X$ rotation per qubit followed by a ring of CNOTs connecting qubit~$i$ to qubit~$(i{+}1)\bmod n$~\cite{kandala_hardware-efficient_2017}. Its shallow depth and single rotation axis trade expressibility for reduced susceptibility to barren plateaus.

\subsubsection{TwoDesign (Simplified Two Design)}



CZ gates are applied in an alternating brick-layer pattern, first on even-indexed pairs and then on odd-indexed pairs, followed by two trainable $R_Y$ rotations per qubit~\cite{cerezo_cost_2021}. This structure forms an approximate unitary 2-design, and is the only circuit in the pool that uses CZ rather than CNOT for entanglement.

\subsubsection{Random (RandomLayers)}



This circuit generates a random gate wiring pattern from a deterministic seed at initialization and frozen. Only the rotation angles within this fixed topology are trained~\cite{boixo_characterizing_2018}. Each layer has an independently randomized connectivity, providing inherent architectural stochasticity that complements the deterministic circuits.

\subsubsection{QAOA (QAOAEmbedding)}



This circuit is adapted from the Quantum Approximate Optimization Algorithm~\cite{farhi_quantum_2014}. It alternates $R_X$ rotations (mixer layer) with pairwise $\mathrm{MultiRZ}$ gates that implement Ising-type $ZZ$ interactions (cost layer). Data re-uploading is built into the template's internal structure, so no external re-uploading loop is needed.

\subsubsection{TTN (Tree Tensor Network)}



In TTN, qubits are partitioned into pairs, and a two-qubit block ($R_Y$--CNOT--$R_Y$ on each qubit) is applied to each pair. Then the outputs are recursively paired and contracted at the next level~\cite{huggins_towards_2019, stoudenmire_supervised_2016}. This hierarchical binary-tree contraction mirrors coarse-graining in condensed-matter tensor networks. Input data is encoded once at the first layer only.

\subsubsection{MPS (Matrix Product State)}



MPS uses the same two-qubit block as TTN but applies the block in a strictly left-to-right sequential sweep along the qubit chain~\cite{stoudenmire_supervised_2016, huggins_towards_2019}. The linear contraction path is shallower in two-qubit gate depth than TTN but limits long-range entanglement, since correlations between distant qubits must propagate through all intermediate sites. Like TTN, input data is encoded once.

\subsubsection{IQP (IQPEmbedding)}

IQP combines a fixed Instantaneous Quantum Polynomial feature map, which applies Hadamard gates, diagonal $R_Z$ rotations encoding both linear~$R_Z(x_i)$ and pairwise-product~$R_Z(x_i \cdot x_j)$ terms, then Hadamard gates again. On top of this fixed encoding, a trainable StronglyEntangling variational layer block with weight shape $(L, n, 3)$ learns to rotate the embedded quantum state into a basis where class separation is maximized~\cite{shepherd_temporally_2009, havlicek_supervised_2019}. The diagonal IQP structure is the only non-trainable encoding in the pool, which makes this circuit the closest analogue to a quantum kernel method.

\subsection{Unified Classification Architecture}





To make quantum classifiers comparable across architectures, every VQC sits inside the same hybrid pipeline:
\begin{equation}
    \hat{y} = \mathbf{W}_{\text{out}} \,
    \tanh\!\Bigl(\text{VQC}\bigl(\tanh(\mathbf{W}_{\text{in}} \mathbf{x}), \boldsymbol{\theta}\bigr)\Bigr).
\end{equation}
The input projection $\mathbf{W}_{\text{in}} \in \mathbb{R}^{n \times d}$ maps a $d$-dimensional feature vector into an $n$-qubit representation, the VQC returns an $n$-dimensional measurement vector, and the output projection $\mathbf{W}_{\text{out}} \in \mathbb{R}^{c \times n}$ produces $c = 2$ class logits. The two projections are wrapped in $\tanh$ to keep the values bounded. We train all circuits with Adam and cross-entropy loss, and use the softmax probability of the positive class as the score function
\begin{equation}
    s_A : \mathcal{D} \to \mathbb{R}
\end{equation}
for each model $A$ in the CFA framework.

\section{Classical baseline models}

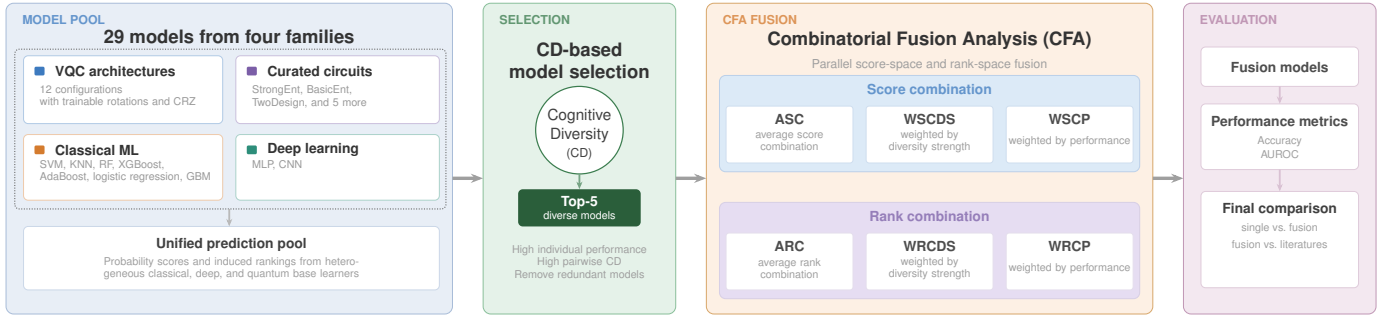
\begin{figure*}[htbp]
\begingroup
\centering
\resizebox{\textwidth}{!}{%
\begin{tikzpicture}[
    >=Stealth,
    every node/.style={font=\sffamily\normalsize},
    arr/.style={-{Stealth[length=5pt,width=4pt]}, color=#1, line width=1.2pt},
    arr/.default={gray!55},
]

\definecolor{poolbg}{HTML}{E8EEF6}
\definecolor{poolbord}{HTML}{A8BCD8}
\definecolor{selbg}{HTML}{E5F2E9}
\definecolor{selbord}{HTML}{89BF9A}
\definecolor{cfabg}{HTML}{FDF0E4}
\definecolor{cfabord}{HTML}{D8A76A}
\definecolor{evalbg}{HTML}{F3E8F0}
\definecolor{evalbord}{HTML}{C097B5}
\definecolor{cblue}{HTML}{3D7ABF}
\definecolor{cviolet}{HTML}{7B5EA7}
\definecolor{corange}{HTML}{D47B2E}
\definecolor{cteal}{HTML}{2D8E7A}
\definecolor{cgreen}{HTML}{3B7D4F}
\definecolor{cdarkgreen}{HTML}{2A5E3A}
\definecolor{scorefill}{HTML}{DCEAF7}
\definecolor{rankfill}{HTML}{E6DDF2}
\definecolor{methfill}{HTML}{FFFFFF}

\fill[poolbg, rounded corners=5pt] (-0.1, -2.7) rectangle (10.1, 4.4);
\draw[poolbord, line width=0.6pt, rounded corners=5pt]
    (-0.1, -2.7) rectangle (10.1, 4.4);
\node[font=\sffamily\footnotesize\bfseries, text=cblue!80, anchor=north west]
    at (0.15, 4.25) {MODEL POOL};

\node[font=\sffamily\large\bfseries, text=black!80]
    at (5.0, 3.65) {29 models from four families};

\fill[white, rounded corners=3pt] (0.3, 1.65) rectangle (4.85, 3.2);
\draw[cblue!40, line width=0.5pt, rounded corners=3pt]
    (0.3, 1.65) rectangle (4.85, 3.2);
\fill[cblue, rounded corners=1pt] (0.55, 2.75) rectangle (0.78, 2.95);
\node[font=\sffamily\small\bfseries, anchor=west, text=black!80]
    at (0.9, 2.85) {VQC architectures};
\node[font=\sffamily\scriptsize, text=gray!80, anchor=north west, text width=4.0cm]
    at (0.55, 2.65) {12 configurations\\with trainable rotations and CRZ};

\fill[white, rounded corners=3pt] (5.15, 1.65) rectangle (9.8, 3.2);
\draw[cviolet!40, line width=0.5pt, rounded corners=3pt]
    (5.15, 1.65) rectangle (9.8, 3.2);
\fill[cviolet, rounded corners=1pt] (5.4, 2.75) rectangle (5.63, 2.95);
\node[font=\sffamily\small\bfseries, anchor=west, text=black!80]
    at (5.75, 2.85) {Curated circuits};
\node[font=\sffamily\scriptsize, text=gray!80, anchor=north west, text width=4.0cm]
    at (5.4, 2.65) {StrongEnt, BasicEnt,\\TwoDesign, and 5 more};

\fill[white, rounded corners=3pt] (0.3, -0.1) rectangle (4.85, 1.4);
\draw[corange!40, line width=0.5pt, rounded corners=3pt]
    (0.3, -0.1) rectangle (4.85, 1.4);
\fill[corange, rounded corners=1pt] (0.55, 0.95) rectangle (0.78, 1.15);
\node[font=\sffamily\small\bfseries, anchor=west, text=black!80]
    at (0.9, 1.05) {Classical ML};
\node[font=\sffamily\scriptsize, text=gray!80, anchor=north west, text width=4.0cm]
    at (0.55, 0.95) {SVM, KNN, RF, XGBoost,\\AdaBoost, logistic regression, GBM};

\fill[white, rounded corners=3pt] (5.15, -0.1) rectangle (9.8, 1.4);
\draw[cteal!40, line width=0.5pt, rounded corners=3pt]
    (5.15, -0.1) rectangle (9.8, 1.4);
\fill[cteal, rounded corners=1pt] (5.4, 0.95) rectangle (5.63, 1.15);
\node[font=\sffamily\small\bfseries, anchor=west, text=black!80]
    at (5.75, 1.05) {Deep learning};
\node[font=\sffamily\scriptsize, text=gray!80, anchor=north west, text width=4.0cm]
    at (5.4, 0.95) {MLP, CNN};

\fill[white, rounded corners=3pt] (0.3, -2.15) rectangle (9.8, -0.70);
\draw[poolbord!60, line width=0.5pt, rounded corners=3pt]
    (0.3, -2.15) rectangle (9.8, -0.70);
\node[font=\sffamily\small\bfseries, text=black!75]
    at (5.05, -1.10) {Unified prediction pool};
\node[font=\sffamily\scriptsize, text=gray!75, text width=8.5cm, align=center]
    at (5.05, -1.65) {Probability scores and induced rankings
    from heterogeneous classical, deep, and quantum base learners};

\draw[black!60, line width=0.7pt, densely dotted, rounded corners=3pt]
    (0.1, -0.3) rectangle (9.95, 3.35);

\draw[arr] (5.0, -0.3) -- (5.0, -0.70);

\fill[selbg, rounded corners=5pt] (10.8, -2.7) rectangle (15.2, 4.4);
\draw[selbord, line width=0.6pt, rounded corners=5pt]
    (10.8, -2.7) rectangle (15.2, 4.4);
\node[font=\sffamily\footnotesize\bfseries, text=cgreen!80, anchor=north west]
    at (11.05, 4.25) {SELECTION};

\node[font=\sffamily\large\bfseries, text=black!80,
      text width=3.6cm, align=center]
    at (13.0, 3.1) {CD-based model selection};

\fill[white] (13.0, 1.5) circle (1.0);
\draw[cgreen, line width=0.8pt] (13.0, 1.5) circle (1.0);
\node[font=\sffamily\normalsize, align=center, text=black!75]
    at (13.0, 1.4) {Cognitive\\Diversity\\{\footnotesize(CD)}};

\fill[cdarkgreen, rounded corners=3pt] (11.6, -0.7) rectangle (14.4, 0.15);
\node[font=\sffamily\small\bfseries, text=white]
    at (13.0, -0.15) {Top-5};
\node[font=\sffamily\scriptsize, text=white!85]
    at (13.0, -0.45) {diverse models};

\draw[arr=cgreen!70] (13.0, 0.5) -- (13.0, 0.15);

\node[font=\sffamily\scriptsize, text=gray!75, text width=3.4cm, align=center]
    at (13.0, -1.5) {High individual performance\\High pairwise CD\\Remove redundant models};

\fill[cfabg, rounded corners=5pt] (15.9, -2.7) rectangle (26.1, 4.4);
\draw[cfabord, line width=0.6pt, rounded corners=5pt]
    (15.9, -2.7) rectangle (26.1, 4.4);
\node[font=\sffamily\footnotesize\bfseries, text=corange!80, anchor=north west]
    at (16.15, 4.25) {CFA FUSION};

\node[font=\sffamily\large\bfseries, text=black!80]
    at (21.0, 3.55) {Combinatorial Fusion Analysis (CFA)};
\node[font=\sffamily\footnotesize, text=gray!75]
    at (21.0, 3.00) {Parallel score-space and rank-space fusion};

\fill[scorefill, rounded corners=4pt] (16.2, 0.55) rectangle (25.8, 2.75);
\draw[cblue!35, line width=0.4pt, rounded corners=4pt]
    (16.2, 0.55) rectangle (25.8, 2.75);
\node[font=\sffamily\small\bfseries, text=cblue!80]
    at (21.0, 2.45) {Score combination};

\foreach \x/\lab/\sub in {
    17.8/ASC/{average score combination},
    21.0/WSCDS/{weighted by diversity strength},
    24.2/WSCP/{weighted by performance}}{
  \fill[methfill, rounded corners=2pt] (\x-1.45, 0.75) rectangle (\x+1.45, 2.05);
  \draw[cblue!25, line width=0.3pt, rounded corners=2pt]
      (\x-1.45, 0.75) rectangle (\x+1.45, 2.05);
  \node[font=\sffamily\small\bfseries, text=black!70] at (\x, 1.75) {\lab};
  \node[font=\sffamily\scriptsize, text=gray!75, text width=2.8cm, align=center] at (\x, 1.25) {\sub};
}

\fill[rankfill, rounded corners=4pt] (16.2, -2.35) rectangle (25.8, -0.15);
\draw[cviolet!30, line width=0.4pt, rounded corners=4pt]
    (16.2, -2.35) rectangle (25.8, -0.15);
\node[font=\sffamily\small\bfseries, text=cviolet!80]
    at (21.0, -0.45) {Rank combination};

\foreach \x/\lab/\sub in {
    17.8/ARC/{average rank combination},
    21.0/WRCDS/{weighted by diversity strength},
    24.2/WRCP/{weighted by performance}}{
  \fill[methfill, rounded corners=2pt] (\x-1.45, -2.15) rectangle (\x+1.45, -0.85);
  \draw[cviolet!20, line width=0.3pt, rounded corners=2pt]
      (\x-1.45, -2.15) rectangle (\x+1.45, -0.85);
  \node[font=\sffamily\small\bfseries, text=black!70] at (\x, -1.15) {\lab};
  \node[font=\sffamily\scriptsize, text=gray!75, text width=2.8cm, align=center] at (\x, -1.65) {\sub};
}

\fill[evalbg, rounded corners=5pt] (26.8, -2.7) rectangle (31.2, 4.4);
\draw[evalbord, line width=0.6pt, rounded corners=5pt]
    (26.8, -2.7) rectangle (31.2, 4.4);
\node[font=\sffamily\footnotesize\bfseries, text=evalbord, anchor=north west]
    at (27.05, 4.25) {EVALUATION};

\fill[white, rounded corners=3pt] (27.2, 2.5) rectangle (30.8, 3.4);
\draw[evalbord!50, line width=0.4pt, rounded corners=3pt]
    (27.2, 2.5) rectangle (30.8, 3.4);
\node[font=\sffamily\small\bfseries, text=black!70] at (29.0, 2.95) {Fusion models};

\fill[white, rounded corners=3pt] (27.2, 0.6) rectangle (30.8, 2.1);
\draw[evalbord!50, line width=0.4pt, rounded corners=3pt]
    (27.2, 0.6) rectangle (30.8, 2.1);
\node[font=\sffamily\small\bfseries, text=black!70]
    at (29.0, 1.7) {Performance metrics};
\node[font=\sffamily\scriptsize, text=gray!75] at (29.0, 1.25) {Accuracy};
\node[font=\sffamily\scriptsize, text=gray!75] at (29.0, 0.95) {AUROC};

\fill[white, rounded corners=3pt] (27.2, -2.1) rectangle (30.8, 0.1);
\draw[evalbord!50, line width=0.4pt, rounded corners=3pt]
    (27.2, -2.1) rectangle (30.8, 0.1);
\node[font=\sffamily\small\bfseries, text=black!75]
    at (29.0, -0.3) {Final comparison};
\node[font=\sffamily\scriptsize, text=gray!75]
    at (29.0, -0.75) {single vs.\ fusion};
\node[font=\sffamily\scriptsize, text=gray!75]
    at (29.0, -1.1) {fusion vs.\ literatures};

\draw[arr=evalbord!60] (29.0, 2.5) -- (29.0, 2.1);
\draw[arr=evalbord!60] (29.0, 0.6) -- (29.0, 0.1);

\draw[-{Stealth[length=8pt,width=6pt]}, gray!75, line width=2.2pt] (10.1, 0.4) -- (10.8, 0.4);
\draw[-{Stealth[length=8pt,width=6pt]}, gray!75, line width=2.2pt] (15.2, 0.4) -- (15.9, 0.4);
\draw[-{Stealth[length=8pt,width=6pt]}, gray!75, line width=2.2pt] (26.1, 0.4) -- (26.8, 0.4);

\end{tikzpicture}}
\endgroup
\caption{\textbf{Workflow of combining quantum and classical models using CFA.}
}
\label{fig:cfa_workflow}
\end{figure*}

This section explains the nine classical models that are added to the 20 quantum classifiers, 7 of which are standard machine learning estimators and 2 are deep learning neural network models. 

\subsection{Classical machine learning models}


We use seven classical machine learning models from scikit-learn. The score function $s_A$ for each model is its probability estimate for class prediction, or a sigmoid of the decision function if such probability estimate is not available. 

Adding classical models to the framework serves two clear purposes. On one hand, they provide strong baselines, which let us measure how well the quantum circuits perform on classification tasks. On the other hand, they increase diversity among the model pools. The models rely on different mechanisms, which range from kernel methods, boosted trees, multilayer perceptrons, convolutional networks, to quantum circuits. Because they process data in different ways, they produce structurally different rank-score functions. That variety strengthens the overall CFA ensemble.

We used the following classical machine learning models as part of the quantum fusion framework: 

\begin{itemize}
    \item Random Forest: An ensemble of 100 decision trees trained on bootstrap samples with random feature subsets. The predicted probability is the average fraction of positive votes across all trees.
    \item Gradient Boosting Machine (GBM): A sequence of 100 depth-3 trees. Each tree fits the negative gradient of the loss from the previous step. Learning rate is 0.1. 
    \item SVM with RBF kernel: Uses the kernel $k(\mathbf{x}, \mathbf{x}') = \exp(-\gamma \|\mathbf{x}-\mathbf{x}'\|^2)$ We apply Platt scaling to obtain probability estimates. The parameters we used are $C=1.0$, $\gamma=\texttt{scale}$.
    \item Logistic Regression: L2-regularized linear classifier with the `lbfgs` solver and `max\_iter=1000`. Serves as a calibrated linear baseline.
    \item K-Nearest Neighbors (KNN): Instance-based classifier with \(k=5\). The predicted probability equals the fraction of positive-class neighbors.
    \item AdaBoost: Adaptive boosting over 100 decision stumps using the SAMME algorithm.
    \item XGBoost: Regularized gradient-boosted trees with second-order loss approximation and column subsampling. Trained for 100 boosting rounds with learning rate 0.3 and max\_depth=6.
\end{itemize}

\subsection{Classical deep learning models}

Both deep learning baselines are implemented in PyTorch and follow the same training setup as the quantum circuits. They are trained with Adam and cross-entropy loss using the same epoch count and batch-size schedule.

The Multi-Layer Perceptron (MLP) is a fully connected model with three hidden layers of 128, 64, and 32 units. Each layer applies a linear transformation followed by ReLU, batch normalization, and dropout with $p = 0.3$. The final layer outputs two logits for binary classification.

The Tabular CNN is a one-dimensional convolutional model that treats the feature vector as a single-channel sequence of shape $(\text{batch}, 1, n_{\text{features}})$. It applies a stack of Conv1d blocks, each consisting of a convolution with kernel size 3 and same padding, followed by ReLU and batch normalization. An 1D adaptive average pooling then collapses the sequence dimension to a fixed-size representation equal to the number of channels in the last convolutional block, regardless of input length. A two-layer fully connected head (Linear $\to$ ReLU $\to$ Dropout($p=0.3$) $\to$ Linear(2)) produces the final logits.

In the hybrid benchmark, the number of convolutional channels and the FC hidden dimension are parameter-matched to the number of circuit qubits, so that the DL model operates at a comparable parameter scale to the VQC circuits it competes against. With the default of 8 qubits, this gives us a single-channel conv block with 8 filters and an 8-dimensional FC hidden layer.

\vspace{-10pt}

\section{CFA framework}

Combinatorial Fusion Analysis (CFA) provides methods and practices for analyzing combination and fusion of multiple scoring systems \cite{hsu_combinatorial_2006}. In this section, we provide a detailed workflow of CFA, which includes diversity-aware model selection and combination methods, as shown in Figure \ref{fig:cfa_workflow}. 

\subsection{Score function, rank function and rank-score function}

\begin{figure*}[htbp]
    \renewcommand{\thefigure}{5}
    \centering
    \includegraphics[width=0.75\linewidth]{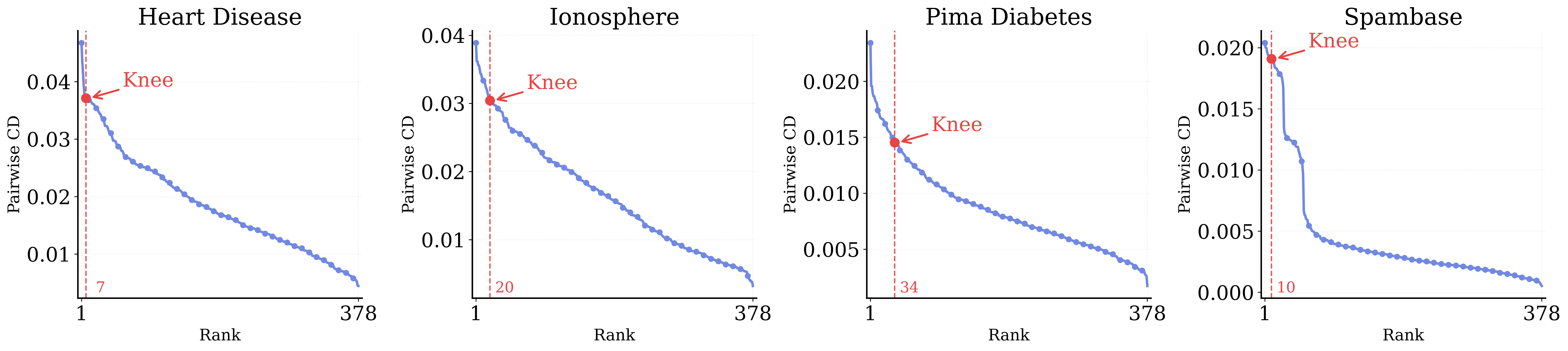}
    \caption{\textbf{Diversity rank-score functions for the four datasets.}}
    \label{fig:pairwise_cd}
\end{figure*}

\begin{figure}[htbp]
    \renewcommand{\thefigure}{4}
    \centering
    \resizebox{0.25\linewidth}{!}{%
        \begin{tikzpicture}[every node/.style={outer sep=0pt}]
             \node[] (D) at (-4, 0) {$D$};
        \node[] (N) at (-2, 0) {$N$};
        \node[] (R) at (-2, -2) {$R$};
        \draw[->, -{Stealth}] (D) -- (N) node[midway, above] {\textcolor{blue}{$r_A$}};
        \draw[->, -{Stealth}] (D) -- (R) node[pos=0.3, below=5pt, xshift = 4pt] {\textcolor{red}{$s_A$}};
        \draw[->, -{Stealth}] (N)--(R) node[pos=0.4, left] {\textcolor{green}{$f_A$}};
        \end{tikzpicture}}
  \caption{\textbf{Relationship among score function $s_A$, rank function $r_A$, and rank-score function $f_A$.}}
  \label{fig:triangle}
\end{figure}
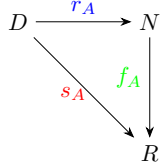

A scoring system is any system that assigns numerical values to data items. It can be a machine learning model producing prediction scores, a quantum circuit outputting expectation values, or clinical measurements such as blood glucose. A scoring system $A$ consists of a score function $s_A$, a rank function $r_A$, and a rank-score function $f_A$, as demonstrated in Figure \ref{fig:triangle}. Formally, Let $D$ be a finite set of data items and let $A$ denote a scoring system defined on $D$. 

The score function of $A$ is a mapping $s_A : D \to \mathbb{R}$, where $s_A(d)$ represents the numerical scores assigned to item $d \in D$. Sorting the data items according to their scores induces a rank function $r_A : D \to \mathbb{N}$, which assigns to each item an integer rank determined by the ordering of $s_A$, with smaller ranks corresponding to higher scores. From these two functions we derive the rank-score function $f_A : \mathbb{N} \to \mathbb{R}$, defined by:
\vspace{-1pt}
\[
f_A(i) = s_A\!\left(r_A^{-1}(i)\right)=(s_A \circ r_A^{-1})(i),
\]

The score function $s_A$ encodes cardinal information in Euclidean space, which maps data items to continuous real-valued numbers in $\mathbb{R}$. The rank function $r_A$ encodes ordinal information in rank space, which maps data items to discrete orders in $\mathbb{N}$. The rank-score function, $f_A$, provides a structured bridge between these two spaces. The RSF is a monotonic non-increasing function in which normalized scores are sorted in descending order using the rank function as the key. The relationship among these three functions is illustrated in Figure \ref{fig:triangle}.
\vspace{-10pt}

\vspace{-10pt}
\subsection{Cognitive diversity and diversity strength}

Cognitive diversity is a diversity measure to quantify the dissimilarity between two scoring systems \cite{hsu_cognitive_2019}. Unlike statistical correlation measures, such as Pearson or Spearman correlation, or computational diversity measures used in ensemble learning, cognitive diversity captures differences between rank-score profiles rather than linear dependence or prediction disagreement alone \cite{hurley_multi-layer_2021}.

Let $A$ and $B$ be two scoring systems defined on the same data set $D$ with rank--score functions $f_A$ and $f_B$. The cognitive diversity between $A$ and $B$ is defined as follows \cite{hsu_cognitive_2019}:

\[
\mathrm{CD}(A,B)
=
\sqrt{
\frac{1}{n}
\sum_{i=1}^{n}
\left(
f_A(i)-f_B(i)
\right)^2
}.
\]
Cognitive diversity is therefore a pairwise measure, and that every pair of scoring systems induces a cognitive diversity value.

Given a collection of scoring systems $\mathcal{S} = \{A_1, \dots, A_m\}$, the diversity strength of a scoring system $A_i$ is defined as the average cognitive diversity between $A_i$ and all other systems,
\[
\mathrm{DS}(A_i)
= \frac{1}{m-1}
\sum_{\substack{j=1 \\ j \ne i}}^{m}
\mathrm{CD}(A_i, A_j).
\]

\subsection{Diversity-aware model selection}

Training a large pool of models and combining all of them rarely works well. Models that perform no better than random guessing weaken the final result, and models that make similar predictions add little beyond redundancy. This experiment uses a diversity-aware selection process that identifies models whose prediction patterns are maximally dissimilar, then clamps the selection to a target size using predictive performance as a tiebreaker.







\begin{figure*}[htbp]
  \centering

  \begin{subfigure}[t]{0.22\textwidth}
    \centering
    \includegraphics[width=\textwidth]{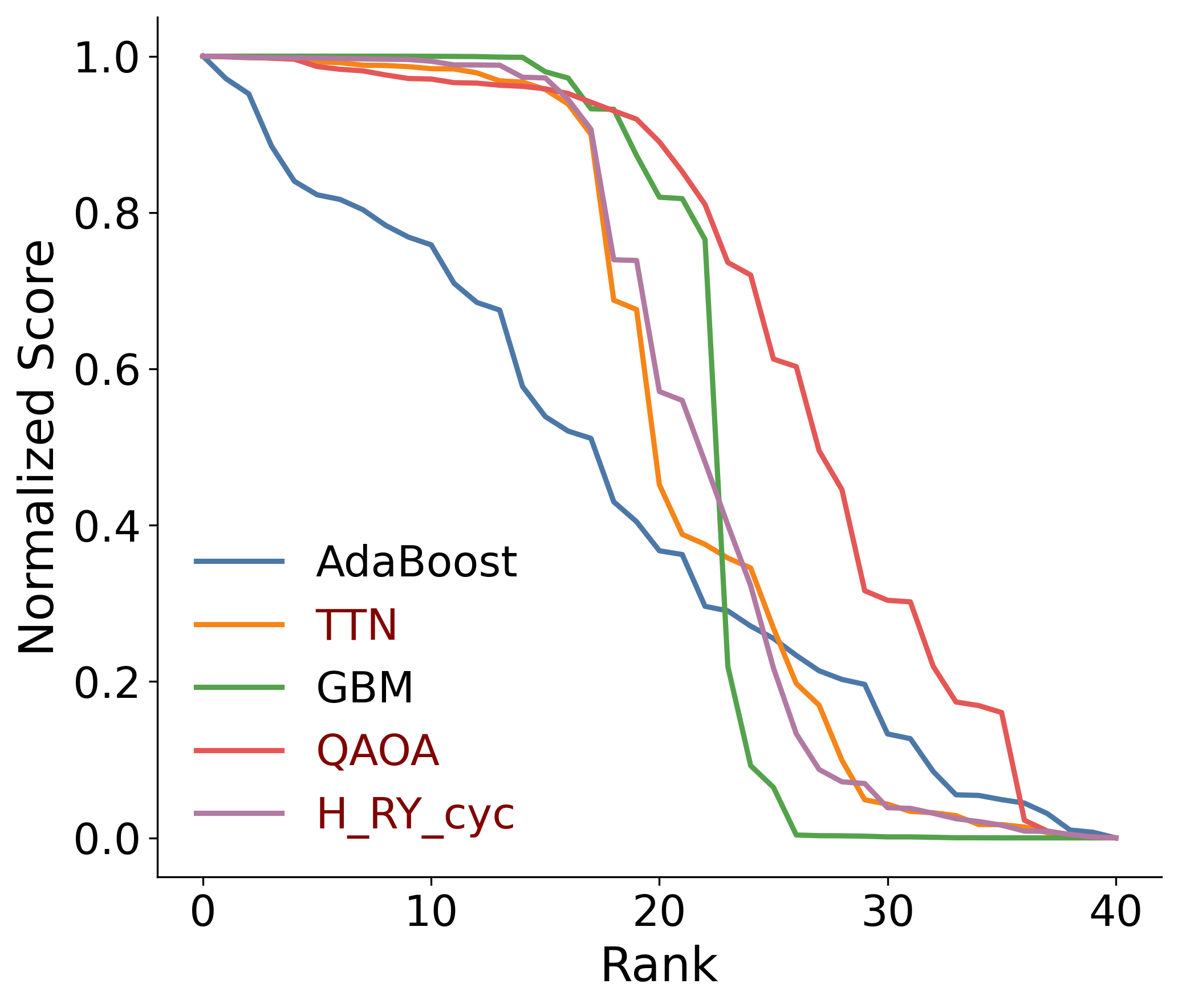}
    \caption{Heart disease}
    \label{fig:rsc_heart}
  \end{subfigure}
  \hfill
  \begin{subfigure}[t]{0.22\textwidth}
    \centering
    \includegraphics[width=\textwidth]{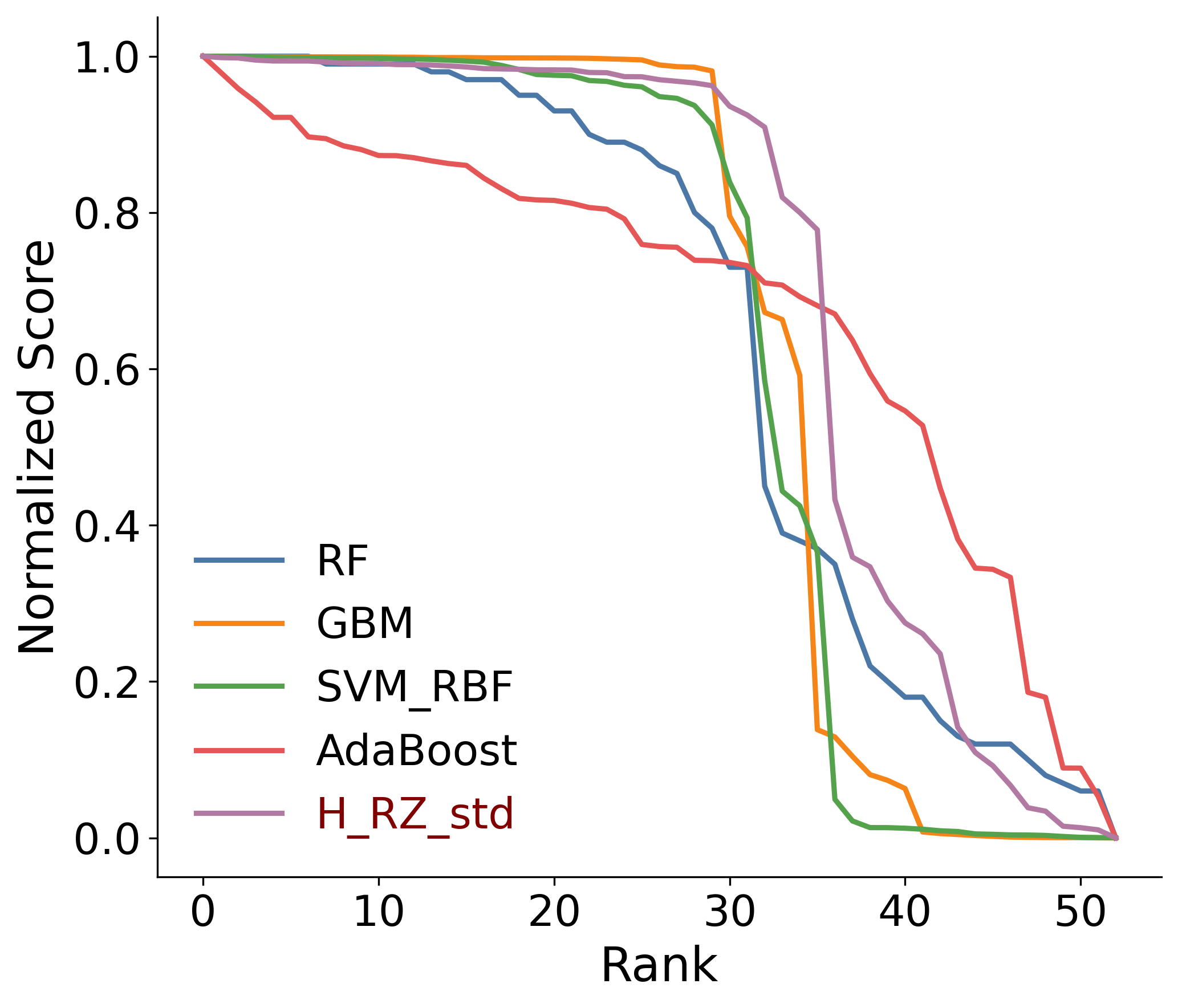}
    \caption{Ionosphere}
    \label{fig:rsc_iono}
  \end{subfigure}
  \hfill
  \begin{subfigure}[t]{0.22\textwidth}
    \centering
    \includegraphics[width=\textwidth]{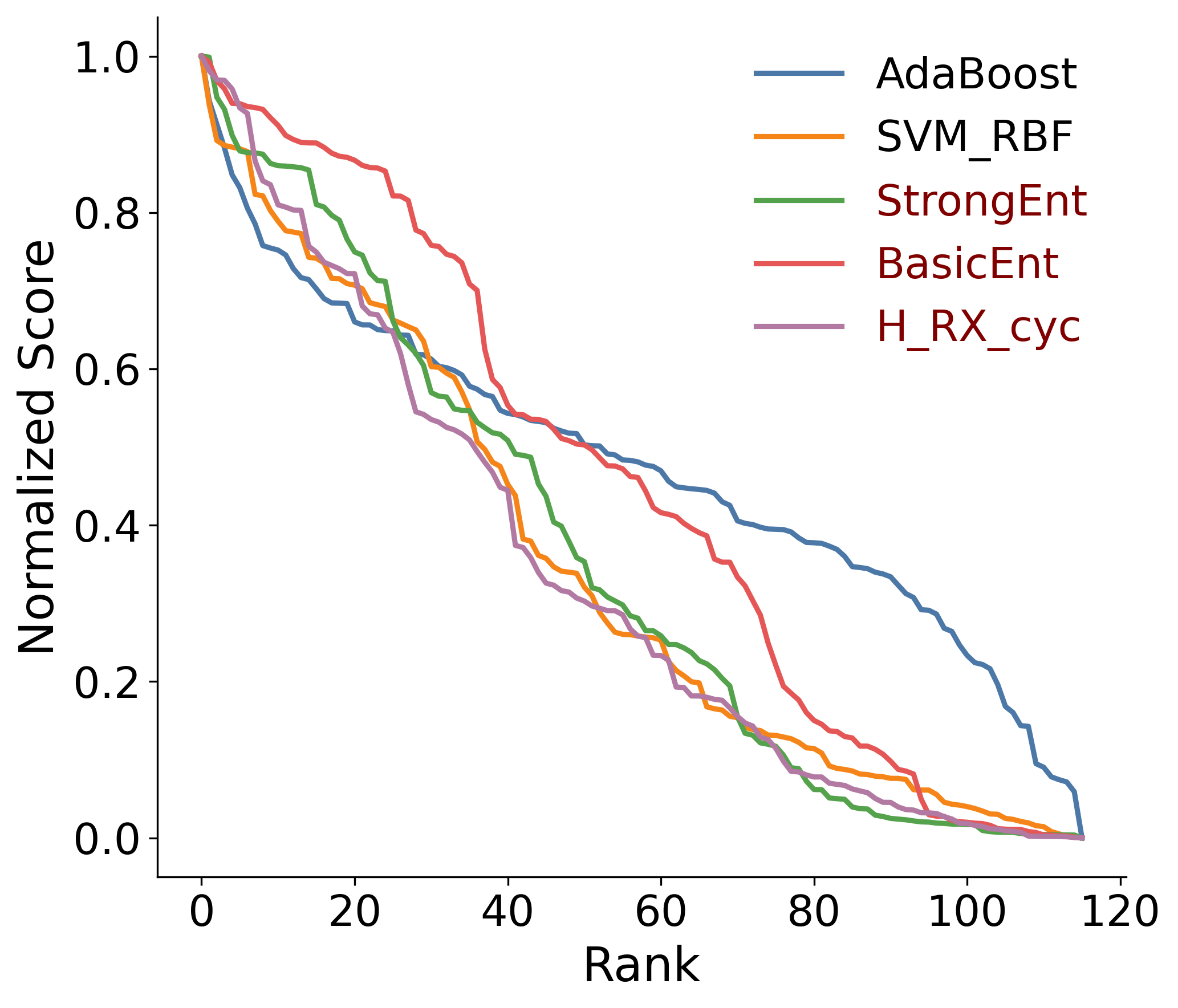}
    \caption{Pima diabetes}
    \label{fig:rsc_pima}
  \end{subfigure}
  \hfill
  \begin{subfigure}[t]{0.22\textwidth}
    \centering
    \includegraphics[width=\textwidth]{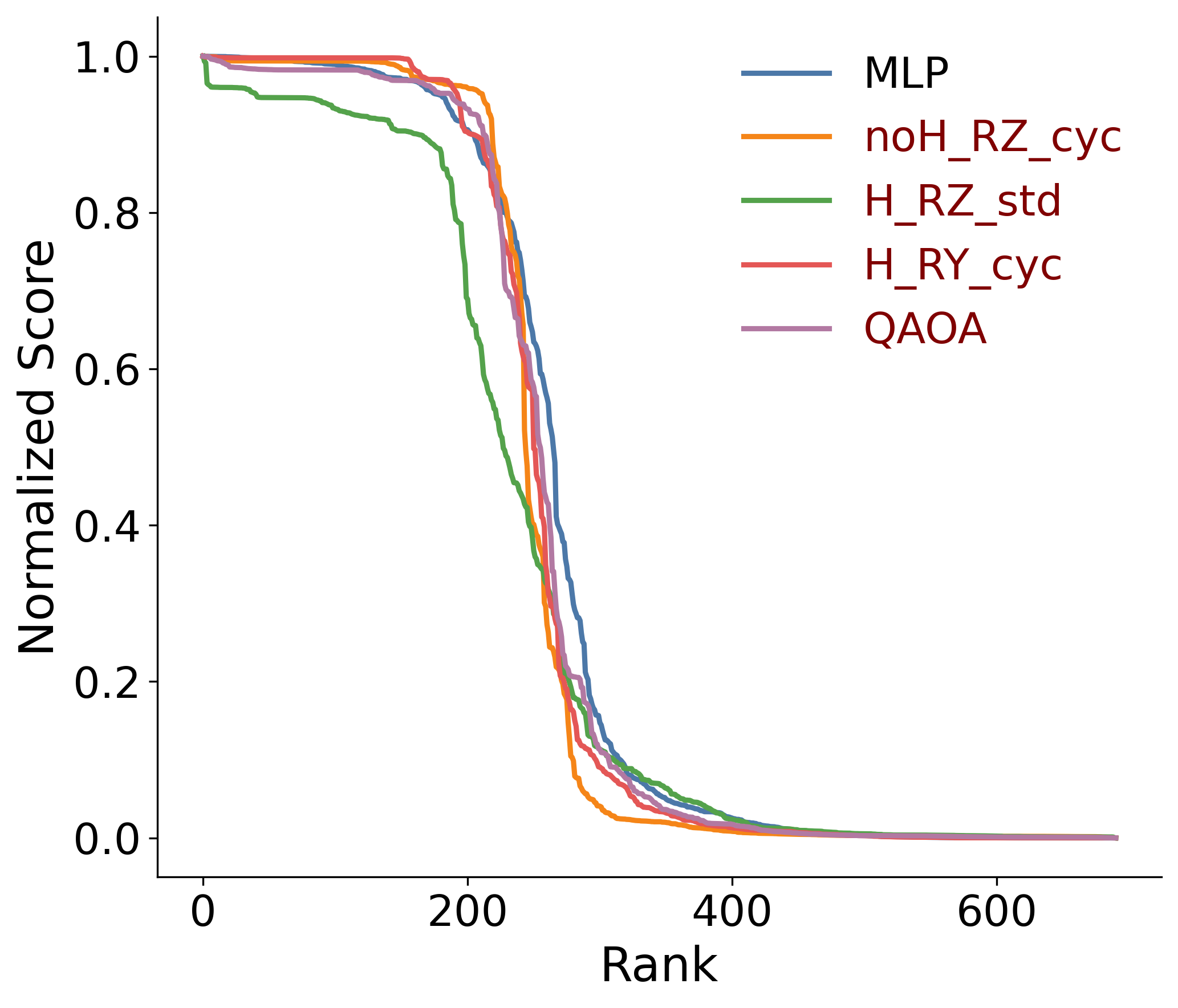}
    \caption{Spambase}
    \label{fig:rsc_spambase}
  \end{subfigure}

  \caption{\textbf{Rank-score function graphs on the test sets of the four datasets} (quantum models are highlighted in maroon).}
  \label{fig:four_panel}
\vskip -0.15in
\end{figure*}

Before selection, we evaluate each model on a held-out validation set and compute its AUROC. Any model with a validation AUROC of 0.5 or lower is removed from the pool. At that threshold, the model performs no better than random guessing in a binary task. This filter acts as a basic quality check. The models that pass form the eligible pool from which the diversity-aware selection chooses candidates for fusion.

Diversity-aware selection method has demonstrated that pairwise diversity between scoring systems can be used to identify complementary subsets for combinatorial fusion \cite{lin_feature_2007}. We adapt pairwise rank-score characteristics analysis to the model selection setting using the cognitive diversity metric.

Given an eligible pool of $t$ scoring systems, we compute the cognitive diversity $CD(A_i, A_j)$ for all $\binom{t}{2}$ unique pairs of models, as shown in Figure \ref{fig:pairwise_cd}. Each CD value quantifies how differently two models rank the data items, a high CD indicates that the two models produce dissimilar prediction orderings, while a low CD indicates redundancy. 

The $\binom{t}{2}$ pairwise CD values, computed on the validation dataset, are sorted in descending order to form a diversity rank-score function curve. This curve plots pair rank (x-axis) against CD value (y-axis). Pairs at the left of the curve represent the most diverse model pairings; as we move to the right, the marginal diversity contribution of additional pairs diminishes. 

To determine where this diminishing return becomes negligible, we apply the Kneedle algorithm, which detects the knee point in a convex, monotonically decreasing curve \cite{satopaa_finding_2011}. The knee is defined as the point of maximum curvature, in our case, where the curve transitions from a steep decline to a flat tail. Formally, Kneedle normalizes the curve to the unit square, computes the difference between the curve and the diagonal connecting its endpoints, and identifies the index of maximum deviation.

All the points in Figure \ref{fig:pairwise_cd} is a pairise CD value and the Knee index clearly cuts off at what the most diverse pairs are. We then obtain all model pairs at or above the knee index. We then extract the unique set of models that appear in at least one of these high-diversity pairs. This gives us a candidate set whose members are, by design, involved in the most diverse pairings in the pool.

The candidate set is then clamped to a target size $N$ (default $N=5$):
\begin{itemize}
    \item If the candidate set contains more than $N$ models, we rank them by AUROC and keep the top $N$, retaining the strongest performers among the diverse candidates.
    \item If the candidate set contains fewer than $N$ models, we backfill from the remaining pool by AUROC, adding the highest-performing unselected models until the target is reached.
    \item If the candidate set contains exactly $N$ models, no adjustment is needed.
\end{itemize}

This approach is category-agnostic, since it does not distinguish between quantum circuits and classical models. Selection is driven entirely by pairwise diversity and individual performance, which allows the algorithm to objectively determine which combination of model types contributes most to the combinations.

\subsection{Combination framework}

CFA characterizes each model by its raw scores (Euclidean space) and the induced ordering of items (rank space), thereby offering two parallel fusion strategies: score combination and rank combination. Within each strategy, model weights can be assigned in three different ways: averaging, weighting by diversity strength, and weighting by performance. Therefore, the CFA framework yields (score combination, rank combination) $\times$ (average weighting, weighted by diversity, weighted by performance) $=6$ different combination approaches by considering (i) whether fusion is performed in the score space or the rank space, and (ii) which weighting scheme is applied during aggregation. The corresponding formulas of these six types of combinations are given below \cite{jiang_enhancing_2023}.

\begin{table*}[h]
\caption{\textbf{Best single-model and best CFA-based combined-model results on four benchmark datasets.} For each dataset, we report the strongest individual model and its performance, together with the best-performing combined model selected by the proposed CFA framework, the corresponding combination type, and the relative improvement over the best single model. Maroon text denotes quantum models.}
\label{tab:comparison_summary}
\centering
\small
\setlength{\tabcolsep}{3pt}
\renewcommand{\arraystretch}{1.1}
\begin{tabularx}{\textwidth}{L{2.1cm} L{2.4cm} L{2.2cm}L{3.8cm} L{1.8cm} L{2cm} l}
\toprule
\multirow{2}{*}{\textbf{Datasets}} & \multicolumn{2}{c}{\textbf{Best single model}} & \multicolumn{4}{c}{\textbf{Best combined model}} \\ \cmidrule(lr){2-3} \cmidrule(lr){4-7}
 & \textbf{Model} & \textbf{Performance} & \textbf{Model} & \textbf{Comb. type} & \textbf{Performance} & \textbf{Improvement} \\ \midrule
 
\multirow{2}{*}{Heart disease} & \multirow{2}{*}{\textcolor{brightmaroon2}{TTN}} & \multirow{2}{*}{AUC: 0.9106} & AdaBoost+\textcolor{brightmaroon2}{TTN}+  & \multirow{2}{*}{ASC/WSCP}  & \multirow{2}{*}{AUC: 0.9444} & \multirow{2}{*}{3.38\%} \\
 &  &   & \textcolor{brightmaroon2}{H\_RY\_cyc}+\textcolor{brightmaroon2}{QAOA} &  &  &\\ \hline
 
\multirow{2}{*}{Ionosphere} 
& GBM or
& \multirow{2}{*}{Acc: 0.9434} 
& \textcolor{brightmaroon2}{H\_RZ\_std}+GBM+
& \multirow{2}{*}{WSCDS} 
& \multirow{2}{*}{Acc: 0.9811} 
& \multirow{2}{*}{3.77\%} \\
& SVM\_RBF &  & SVM\_RBF+AdaBoost &  &  &  \\ \hline

\multirow{2}{*}{Pima diabetes} & Adaboost or  & \multirow{2}{*}{Acc: 0.7845} & AdaBoost+\textcolor{brightmaroon2}{StrongEnt}  & \multirow{2}{*}{ASC/WSCP} & \multirow{2}{*}{Acc: 0.8448} & \multirow{2}{*}{6.03\%} \\
 & SVM\_RBF or \textcolor{brightmaroon2}{StrongEnt} & & +\textcolor{brightmaroon2}{BasicEnt}  &   &  & \\ \hline

\multirow{2}{*}{Spambase} 
& \multirow{2}{*}{MLP} & \multirow{2}{*}{Acc: 0.9349}  
& \multirow{2}{*}{MLP+\textcolor{brightmaroon2}{noH\_RZ\_cyc}} 
& ASC/WSCDS/
& \multirow{2}{*}{Acc: 0.9407} 
& \multirow{2}{*}{0.58\%} \\
 &  &  &  &  WSCP& &\\ 
 \bottomrule
\end{tabularx}
\end{table*}

{\setlength{\abovedisplayskip}{3pt}\setlength{\belowdisplayskip}{3pt}\setlength{\abovedisplayshortskip}{1pt}\setlength{\belowdisplayshortskip}{1pt}
\textbf{Average Score Combination (ASC):}
\[
s_{\mathrm{ASC}}(d_i)
=
\frac{1}{m}\sum_{j=1}^{m} s_{A_j}(d_i).
\]

\textbf{Average Rank Combination (ARC):}
\[
s_{\mathrm{ARC}}(d_i)
=
\frac{1}{m}\sum_{j=1}^{m} r_{A_j}(d_i).
\]

\textbf{Weighted Score Combination by Diversity Strength (WSCDS):}
\[
s_{\mathrm{WSCDS}}(d_i)
=
\sum_{j=1}^{m}
\frac{\mathrm{DS}(A_j)}
{\sum_{k=1}^{m}\mathrm{DS}(A_k)}
\, s_{A_j}(d_i).
\]

\textbf{Weighted Rank Combination by Diversity Strength (WRCDS):}
\[
s_{\mathrm{WRCDS}}(d_i)
=
\sum_{j=1}^{m}
\frac{1/\mathrm{DS}(A_j)}
{\sum_{k=1}^{m} 1/\mathrm{DS}(A_k)}
\, r_{A_j}(d_i).
\]

\textbf{Weighted Score Combination by Performance (WSCP):}
\[
s_{\mathrm{WSCP}}(d_i)
=
\sum_{j=1}^{m}
\frac{P(A_j)}
{\sum_{k=1}^{m} P(A_k)}
\, s_{A_j}(d_i).
\]

\textbf{Weighted Rank Combination by Performance (WRCP):}
\[
s_{\mathrm{WRCP}}(d_i)
=
\sum_{j=1}^{m}
\frac{1/P(A_j)}
{\sum_{k=1}^{m} 1/P(A_k)}
\, r_{A_j}(d_i).
\]
}

Figure \ref{fig:cfa_workflow} presents the overall workflow for combining quantum and classical models using CFA. The framework consists of four main steps. Firstly, base models are trained from four different model families: VQC architectures, curated quantum circuits, classical machine learning models, and deep learning models. Their predictions are obtained on the test set. Secondly, a model selection step is applied, where the top five models are chosen based on cognitive diversity and individual performance in order to reduce homogeneous selection in terms of each model's rank-score characteristics. Cognitive diversity and individual performances are computed on the split validation set. Thirdly, these selected models are then combined parallelly through CFA using raw score combination and rank combination. Within each strategy, three weighting schemes are considered: average weighting, weighting by diversity strength, and weighting by performance. Finally, the combined models are evaluated using two common performance metrics, Accuracy and AUROC, as these two are the commonly preferred metrics for classification problems. The combined models' performance is compared with the five selected base models as well as recent results reported in the literature.

\vspace{-4pt}

\section{Results}

The rank--score function graphs for the four datasets are shown in Figure \ref{fig:four_panel}. In these plots, the x-axis represents the rank, and the y-axis represents the normalized score. 
Each curve corresponds to one model. The five models from each dataset are selected through the model selection method mentioned in Section VI-C. Across the four datasets, we observe noticeable differences in curve shapes between models, especially between the VQCs and classical models. These differences reflect distinct scoring behaviors, which in the CFA framework correspond to cognitive diversity. Such diversity is essential for effective fusion.
\vspace{-5pt}
\begin{table}[htbp]
\centering
\caption{\textbf{List of selected models per dataset}}
\label{tab:selected_models}
\begin{tabular}{l l l}
\toprule
\textbf{Dataset} & \textbf{Category} & \textbf{Models} \\
\midrule
\multirow{2}{*}{Heart disease} & Quantum & \textcolor{brightmaroon2}{TTN, QAOA, H\_RY\_cyc}\\
                                & Classical & AdaBoost, GBM \\
\midrule
\multirow{2}{*}{Ionosphere} & Quantum & \textcolor{brightmaroon2}{H\_RZ\_std} \\
            & Classical & RF, GBM, SVM\_RBF, AdaBoost \\
\midrule
\multirow{2}{*}{Pima diabetes} & Quantum & \textcolor{brightmaroon2}{StrongEnt, BasicEnt, H\_RX\_cyc} \\
            & Classical & AdaBoost, SVM\_RBF \\
\midrule
\multirow{2}{*}{Spambase} & Quantum & \textcolor{brightmaroon2}{\makecell[tl]{noH\_RZ\_cyc, H\_RZ\_std, H\_RY\_cyc, \\ QAOA}} \\
            & Classical & MLP \\
\bottomrule
\end{tabular}
\vskip -0.1in
\end{table}

The subsets in Table~\ref{tab:selected_models} vary substantially across the four datasets due to category-agnostic and diversity-aware selection rule. At least one quantum circuit is selected on every dataset and not a single model is picked for all datasets. The quantum mix itself shifts from a tensor-network circuit on Heart disease, to hardware-efficient circuits on Pima diabetes, to the Ising-inspired QAOA circuit on Spambase. The pipeline is not biased toward classical models nor toward any one ansatz family on the quantum side. 

\begin{table*}[h]
\centering
\caption{\textbf{Benchmark comparison of the proposed method vs. recent reported results on four datasets.} For each dataset, three studies are listed with publication year, model, and reported performance. The final row for each dataset shows the proposed classical-quantum combined model and its performance. Maroon text denotes quantum circuits from our framework.}
\label{tab:benchmark_summary}
\begin{tabular}{p{18mm}l >{\centering\arraybackslash}p{10mm}>{\raggedright\arraybackslash}p{60mm}c}
\toprule
\textbf{Dataset} & \textbf{Paper} & \textbf{Year} & \textbf{Model} & \textbf{Performance} \\
\hline

\multirow{4}{*}{Heart disease}& Qi et al. \cite{zhenya_hybrid_2021} & 2021& Cost-sensitive ensemble (RF+LR+SVM+ELM+KNN, full features)& AUC: 0.8799\\
& Mukherjee et al. \cite{mukherjee_diagnosis_2017}& 2017& SVM& AUC: 0.8985\\
& Heidari et al. \cite{heidari_heart_2024}& 2022& Hybrid quantum neural network and random forest (HQRF)& AUC: 0.9164\\
& \textbf{Our method (CFA)}& 2026& \textbf{AdaBoost+\textcolor{brightmaroon2}{TTN}+\textcolor{brightmaroon2}{H\_RY\_cyc}+\textcolor{brightmaroon2}{QAOA}}& AUC: \textbf{0.9444}\\
\hline

\multirow{4}{*}{Ionosphere} 
& Santucci \cite{santucci_quantum_2017} & 2017 & Quantum nearest mean classifier (QNMC) & Acc: 0.8350 \\
& Tychola et al. \cite{tychola_quantum_2023} & 2023 & Quantum SVM & Acc: 0.8400 \\
& Li et al. \cite{li_improved_2019} & 2019 & PSOPD + AdaBoost & Acc: 0.9200 \\
& \textbf{Our method (CFA)} & 2026 & \textbf{\textcolor{brightmaroon2}{H\_RZ\_std}+GBM+SVM\_RBF+AdaBoost} & Acc: \textbf{0.9811} \\ 
\hline

\multirow{4}{*}{Pima diabetes}& Azzahra et al. \cite{rizal_type_2025} & 2025& SMOTE + XGB & Acc: 0.7460\\
& Sheoran et al. \cite{sheoran_robust_2025} & 2025 & SMOTE + Quantum SVM & Acc: 0.7550\\
& Rahutomo et al. \cite{rahutomo_comparison_2025} & 2025& Logistic Regression& Acc: 0.7773\\
& \textbf{Our method (CFA)} & 2026 & \textbf{AdaBoost+\textcolor{brightmaroon2}{StrongEnt}+\textcolor{brightmaroon2}{BasicEnt}} & Acc: \textbf{0.8448} \\
\hline

\multirow{4}{*}{Spambase} 
& Loglisci et al. \cite{loglisci_hybrid_2023} & 2023 & Hybrid quantum-classical online learning framework (HYQOL) & Acc: 0.9000 \\
& Rahman et al. \cite{rahman_commentclass_2024} & 2024 & RF + Adaboost + SVM + Soft-Voting & Acc: 0.9229 \\
& Ayo et al. \cite{ayo_hybrid_2024} & 2024 & Fuzzy logic + genetic search + deep learning model & Acc:0.9400 \\
& \textbf{Our method (CFA)} & 2026 & \textbf{MLP+\textcolor{brightmaroon2}{noH\_RZ\_cyc}} & Acc: \textbf{0.9407} \\
\bottomrule

\end{tabular}
\vskip -0.15in
\end{table*}

Since five models are used as base models, the CFA framework can generate 
$6 \times (2^5 - 5 - 1) = 156$ model combinations. Due to space limitations, we report only the best individual base model(s), the best combined model(s), and the performance metrics with highest improvement for each dataset, as shown in Table \ref{tab:comparison_summary}. In this table, we compare the best single model with the best combined model identified by the CFA framework across the four benchmark datasets. 

For the heart disease dataset, performance is evaluated by AUC, while for the other three datasets, accuracy is used. Across all four datasets, the combined models consistently outperform the strongest individual model. These results support the central argument of this work: instead of relying on a single classical or quantum learner, a fusion of classical and quantum models can exploit complementary strengths among heterogeneous models and lead to measurable performance gains. The best single model on the heart disease dataset is a quantum model TTN with an AUC of 0.9106. The best combined model further improves the performance to an AUC of 0.9444, yielding a relative improvement of $3.38\%$ under ASC or WSCP. Notably, the optimal combination includes both the classical model with quantum circuits. Together with the rank-score function graph shown in Figure \ref{fig:four_panel}, this suggests that the classical learner AdaBoost is highly diverse from the quantum models, and that combining these two distinct types of models can lead to performance gains.

For the Ionosphere dataset, combining one quantum model with three classical machine learning models yields a relatively large improvement, with accuracy increasing from 0.9434 to 0.9811. For Pima Diabetes, the improvement is the largest among all four datasets. The best individual model, AdaBoost, SVM\_RBF, or StrongEnt, achieves an accuracy of 0.7845, whereas the best combined model reaches 0.8448 under average score combination or weighted score combination by performance. On Spambase, performance improves more modestly, from 0.9349 to 0.9407, corresponding to a gain of $0.58\%$. The rank–score functions figure for Spambase shows that models are very similar to one another, which helps explain the relatively small improvement.

Overall, the results in Table \ref{tab:comparison_summary} show that the proposed CFA framework can effectively identify beneficial hybrid combinations and suitable fusion rules for different datasets.

Table \ref{tab:benchmark_summary} compares the proposed method with several benchmark results reported in the literature for the same datasets. The baselines include both conventional models and more recent ensemble or quantum-related approaches.

On the heart disease dataset, the proposed method achieves an AUC of 0.9444, which is higher than the compared results from Qi et al. \cite{zhenya_hybrid_2021} (0.8799), Mukherjee et al. \cite{mukherjee_diagnosis_2017} (0.8985), and Heidari et al. \cite{heidari_heart_2024} (0.9164). While some prior studies already used ensemble model and hybrid quantum-classicial models, our result shows that using CFA to combine classical and quantum models can achieve higher performance beyond these existing approaches. 

For Ionosphere, the proposed model reaches an accuracy of 0.9811, substantially outperforming the listed baselines. This margin is quite large, suggesting that the CFA framework is especially effective on this dataset. The result indicates that CFA is able to combine the strengths of multiple models in a way that produces a much stronger predictor than either individual quantum or classical baselines. On Pima diabetes, the proposed model attains an accuracy of 0.8448, exceeding the compared results. For Spambase, our method obtains an accuracy of 0.9407, outperforming all the baselines.

Taken together, the benchmark results demonstrate that the proposed method achieves better performance across all four datasets. Since the compared baselines already include ensemble models and quantum-based methods, these improvements provide strong evidence that the advantage comes not merely from using more models, but from the CFA mechanism itself, which systematically explores and selects effective combinations. The diversity-aware design of CFA enables consistent performance improvement. It also suggests that classical and quantum models contain complementary information, and CFA offers a practical way to exploit that complementarity. 

\vspace{-5pt}

\section{Conclusion}

In this study, we presented a robust diversity-aware fusion framework that bridges the gap between quantum and classical machine learning by using Combinatorial Fusion Analysis (CFA). Our methodology utilized a hybrid pool of quantum and classical machine learning models. We then employed a model selection method to identify the top five most diverse models using Cognitive Diversity (CD). These models were later combined through both score-based and rank-based fusion, using strategies such as uniform averaging and weighting by diversity strength and performance.

Our experimental results across four binary classification benchmarks demonstrate that this CFA-based fusion consistently outperforms the best individual models, as well as recent state-of-the-art ensemble and quantum approaches. Notably, the optimal combinations across the four datasets were composed of both quantum and classical models. This highlights that the diversity between quantum and classical models can be effectively harnessed through the CFA mechanism to achieve performance gains that surpass those of either quantum or classical models alone.

We envision several directions for future research. First, the current single-layer CFA fusion pipeline can be expanded into a Multi-layer Combinatorial Fusion (MCF) framework \cite{hurley_multi-layer_2021, zhong_combining_2019}. By utilizing the expansion-reduction algorithm for iterative model fusion, we may further enhance current performance. Additionally, we plan to evaluate the CFA framework against quantum-induced noise to determine if observed diversity patterns and fusion gains remain robust in practical, noisy environments.

\bibliographystyle{ieeetr}
\bibliography{quantum_fusion.bib}

\end{document}